\documentclass[aps,prl,twocolumn,superscriptaddress,floatfix,amsmath,amsfonts]{revtex4-1}

\usepackage{graphicx}
\usepackage{braket}
\usepackage{bm}
\usepackage{xcolor}
\usepackage{upgreek}
\usepackage{tabularx}
\usepackage[normalem]{ulem}
\usepackage[utf8]{inputenc}
\usepackage{acronym}

\usepackage{hyperref}

\begin{document}
\title{Classical and semiclassical description of Rydberg excitons in cuprous oxide}
\author{Jan Ertl}
\affiliation{Institut für Theoretische Physik 1, Universit\"at
  Stuttgart, 70550 Stuttgart, Germany}
\affiliation{Experimentelle Physik 2, Technische Universit\"at Dortmund,
  44221 Dortmund, Germany}
\author{Patric Rommel}
\affiliation{Institut für Theoretische Physik 1, Universit\"at
  Stuttgart, 70550 Stuttgart, Germany}
\author{Michel Mom}
\affiliation{Institut für Theoretische Physik 1, Universit\"at
  Stuttgart, 70550 Stuttgart, Germany}
\author{J\"org Main}
\email[Email: ]{main@itp1.uni-stuttgart.de}
\affiliation{Institut für Theoretische Physik 1, Universit\"at
  Stuttgart, 70550 Stuttgart, Germany}
\author{Manfred Bayer}
\affiliation{Experimentelle Physik 2, Technische Universit\"at Dortmund,
  44221 Dortmund, Germany}
\date{\today}

\begin{abstract}
Experimental and theoretical investigations of excitons in cuprous
oxide have revealed a significant fine-structure splitting of the
excitonic Rydberg states caused by a strong impact of the valence band
structure.
We provide a semiclassical interpretation of that splitting by
investigating the classical dynamics of the excitonic electron-hole
pair beyond the hydrogen-like model.
Considering the slow motion of Rydberg excitons in coordinate space
compared to the fast dynamics of quasispin and hole spin we use an
adiabatic approach and energy surfaces in momentum space for the
computation of the exciton dynamics.
We observe quasi-periodic motion on near-integrable tori.
Semiclassical torus quantization yields the energy regions of the
fine-structure splitting of $n$-manifolds in agreement with quantum
mechanical computations.
\end{abstract}

\maketitle

% Abbreviations
\acrodef{PSOS}{Poincar\'e surface of section}

%\paragraph{Introduction}
Excitons in a semiconductor like cuprous oxide are created by the
excitation of an electron from the valence band to the conduction
band.
The electron and the remaining hole in the valence band form a
hydrogen-like system, which, however, is influenced by the band
structure of the crystal.
Since the early experiments on cuprous oxide by Gross~\cite{Gross1956}
the experimental techniques have made enormous progress.
In 2014 Kazimierczuk \emph{et al.}\ realized the excitation of
excitons with principal quantum numbers up to $n=25$ for the yellow
series in cuprous oxide~\cite{GiantRydbergExcitons}.
These results have significantly inspired the field of giant Rydberg
excitons.
Due to the influence of the cubic $O_{\mathrm{h}}$ symmetry of the
crystal, the exciton sequence shows deviations from a perfect
hydrogen-like spectrum, viz.\ a fine-structure splitting of the
$n$-manifolds \cite{ObservationHighAngularMomentumExcitons,ImpactValence}.
 
The complex band dispersion causing the fine-structure splitting can
be described by introducing empirical quantum defects similar to the
quantum defects in atoms \cite{Schoene2016}.
The excitons are then described by a modified Rydberg formula, where
the principal quantum number $n$ is replaced by an effective quantum
number $n_{\mathrm{eff}} = n - \delta_{n,l}$.
For the yellow exciton series the quantum defects $\delta_{n,l}$ are
positive and grow with decreasing angular momentum $l$.
They also slightly grow as function of $n$ until saturating for large $n$.
However, neither this phenomenological approach nor exact quantum
computations for excitons provide the basis for an intuitive physical
interpretation of the spectra.
Similar is true for other complex systems, a prominent example being
the hydrogen atom in a magnetic field, which is a quantum system,
whose classical analogue shows a transition from regular to chaotic
dynamics, and whose spectra in the non-perturbative regime can hardly
be explained using a pure quantum mechanical terminology.
Indeed, this system has become a prototype example for the
investigation of \emph{quantum chaos} by means of classical and
semiclassical methods \cite{Fri89}.
For the hydrogen atom semiclassical methods provide a deeper
insight by, e.g., relating the energy spacing of Rydberg states to
the periods of classical Kepler orbits. 
To the best of our knowledge the classical dynamics of excitons and
its relation to excitonic spectra has not yet been investigated.
In particular it is an interesting question whether the influence of
the band structure leads to regular, chaotic, or mixed regular-chaotic
exciton dynamics.

In this Rapid Communication we use an adiabatic approach to compute
the classical orbits of excitons in cuprous oxide including the
complex valence band structure.
We show that the classical exciton dynamics significantly deviates
from the hydrogen-like model, and demonstrate that the fine-structure
splitting of the $n$-manifolds can be interpreted via a semiclassical
torus quantization of the quasi-periodic exciton orbits.

%\paragraph{Theory and results}
As mentioned above, a full description of excitons in cuprous oxide
needs to consider the cubic $O_{\rm h}$ symmetry of the system.
The uppermost valence band of cuprous oxide  belongs to the
irreducible representation $\Gamma_5^+$ \cite{Schoene2016,SchoeneLuttinger}.
To consider this three-fold degenerate valence band one can introduce
a quasispin $I=1$.
Additionally, the hole in the valence band has spin $S_{\rm h}=1/2$.
Quasispin and hole spin are coupled by the spin-orbit term
\begin{equation}
  H_{\mathrm{SO}}=\frac{2}{3}\Delta
  \left(1+\frac{1}{\hbar^2}\boldsymbol{I}\cdot\boldsymbol{S}_{\mathrm{h}}\right)\, ,
\end{equation}
where $\Delta$ is the spin-orbit coupling, and the components of the
vectors $\boldsymbol{I}$ and $\boldsymbol{S}_{\rm h}$ are the three
spin matrices for $I=1$ and $S_{\rm h}=1/2$, respectively.
This leads to a splitting of the $\Gamma_5^+$ band
into a higher lying two-fold degenerate $\Gamma_7^+$ band,
connected to the yellow exciton series, and a lower lying four-fold
degenerate $\Gamma_8^+$ band, connected to the green exciton
series \cite{Schoene2016,SchoeneLuttinger,koster1963properties}.
Introducing relative and center-of-mass coordinates for the electron
and hole and neglecting the center-of-mass momentum, the Hamiltonian
for excitons in cuprous oxide is given by
\cite{ImpactValence,Uihlein1981,Lipari1977}
\begin{equation}
  H = E_{\mathrm{g}} + \frac{\gamma'_1}{2m_0} \boldsymbol{p}^2
  -\frac{e^2}{4\pi\varepsilon_0\varepsilon|\boldsymbol{r}|}
  + H_{\mathrm{b}}(\boldsymbol{p},\boldsymbol{I},\boldsymbol{S}_{\rm h})\, ,
\label{eq:H}
\end{equation}
where the first term is the gap energy between the uppermost valence
band and the lowest conduction band.
The second and third term correspond to the hydrogen-like model with
$\gamma'_1= \gamma_1 + m_0/m_{\mathrm{e}}$ and the screened Coulomb
potential with the dielectric constant $\varepsilon$.
The fourth term
\begin{align}
  &H_{\mathrm{b}} (\boldsymbol{p},\boldsymbol{I},\boldsymbol{S}_{\mathrm{h}})
  = H_{\mathrm{SO}}+\frac{1}{2\hbar^2m_0}
  \big[4\gamma_2\hbar^2\boldsymbol{p}^2\nonumber\\[1ex]
  &-6\gamma_2(p^2_1\boldsymbol{I}^2_1+{\rm c.p.})
    -12\gamma_3(\{p_1,p_2\}\{\boldsymbol{I}_1,\boldsymbol{I}_2\}+{\rm c.p.})\nonumber\\[1ex]
  &-12\eta_2(p^2_1\boldsymbol{I}_1\boldsymbol{S}_{\rm h1}+{\rm c.p.})
    +2(\eta_1+2\eta_2)\boldsymbol{p}^2(\boldsymbol{I}\cdot\boldsymbol{S}_{\rm h})\nonumber\\[1ex]
  &-12\eta_3(\{p_1,p_2\}(\boldsymbol{I}_1\boldsymbol{S}_{\rm h2}
    +\boldsymbol{I}_2\boldsymbol{S}_{\rm h1})+{\rm c.p.})\big] \, ,
\label{eq:Hb}
\end{align}
accounts for the cubic band structure. 
Here, $m_0$ is the free-electron mass, $\{a,b\}=\frac{1}{2}(ab+ba)$
denotes the symmetrized product, c.p.\ stands for cyclic permutation,
and the $\gamma_i$ and $\eta_i$ are the Luttinger parameters.
In our computations we use the same material parameters as
given in Ref.~\cite{Rommel2020}.

The Hamiltonian~\eqref{eq:H} can be treated in a quantum theoretical
framework \cite{ImpactValence}, where the lifting of the degeneracies
of the Rydberg states can be observed.
We want to provide a semiclassical interpretation for these phenomena
by connecting the observed spectra to classical exciton orbits.
Advancing to Rydberg states characterized by high principal quantum
numbers, we make use of the Bohr correspondence principle to replace
the operators for the relative motion by classical variables.
In comparison to the hydrogen-like approach the band structure needs
to be considered via quasispin and hole spin.
With the Rydberg energy for the exciton
$E_{\mathrm{Ryd}}=13.6\,\mathrm{eV}/(\gamma_1'
\varepsilon^2)=0.087\,\mathrm{eV}$, the energy spacing
$E_{\mathrm{Ryd}}/n^2-E_{\mathrm{Ryd}}/(n+1)^2 \approx 2
E_{\mathrm{Ryd}}/n^3$ between adjacent 
Rydberg excitons is small compared to the spin-orbit coupling $\Delta$
already for Rydberg excitons with $n\ge 3$, which means that the
characteristic timescale of the spin dynamics is short compared to
that of the exciton dynamics in coordinate space.
This allows for using an adiabatic approach by assuming that the fast
spin dynamics reacts instantly to a change in the slow relative
motion, in analogy to, e.g., the Born-Oppenheimer approximation for
molecules, where the fast electron motion is assumed to react
instantly to a slow change in the core configuration.
In the following the spin dynamics and the exciton dynamics in
coordinate space are treated quantum mechanically and classically,
respectively.

\begin{figure}
  \includegraphics[width=\columnwidth]{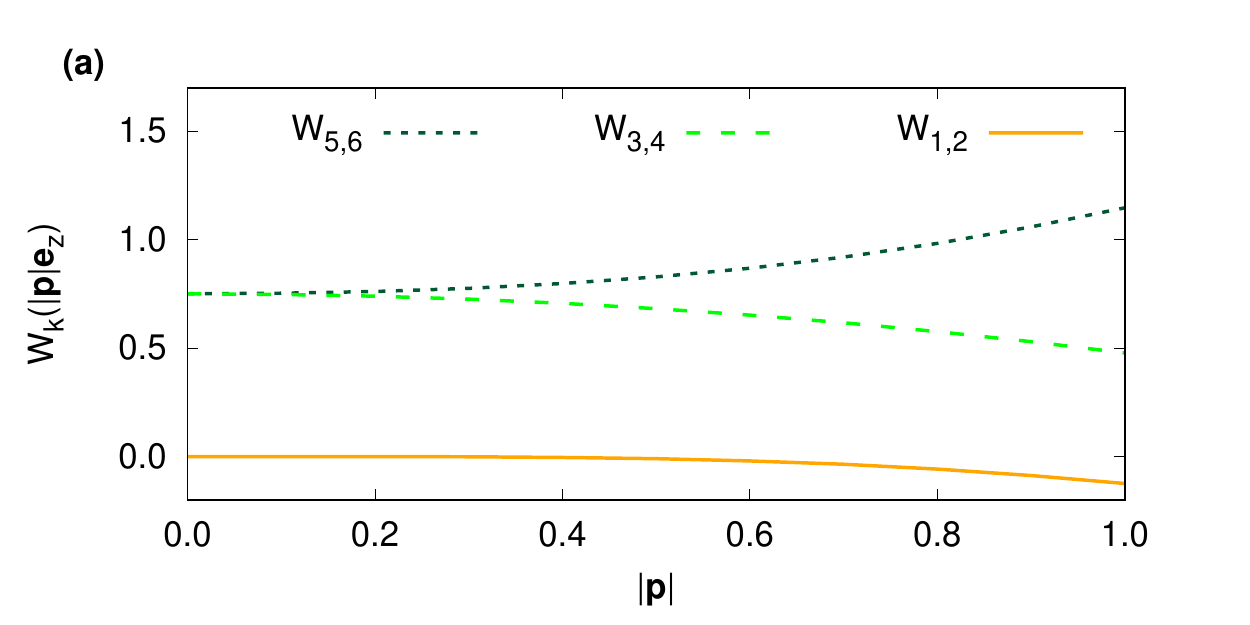}
  \includegraphics[width=\columnwidth]{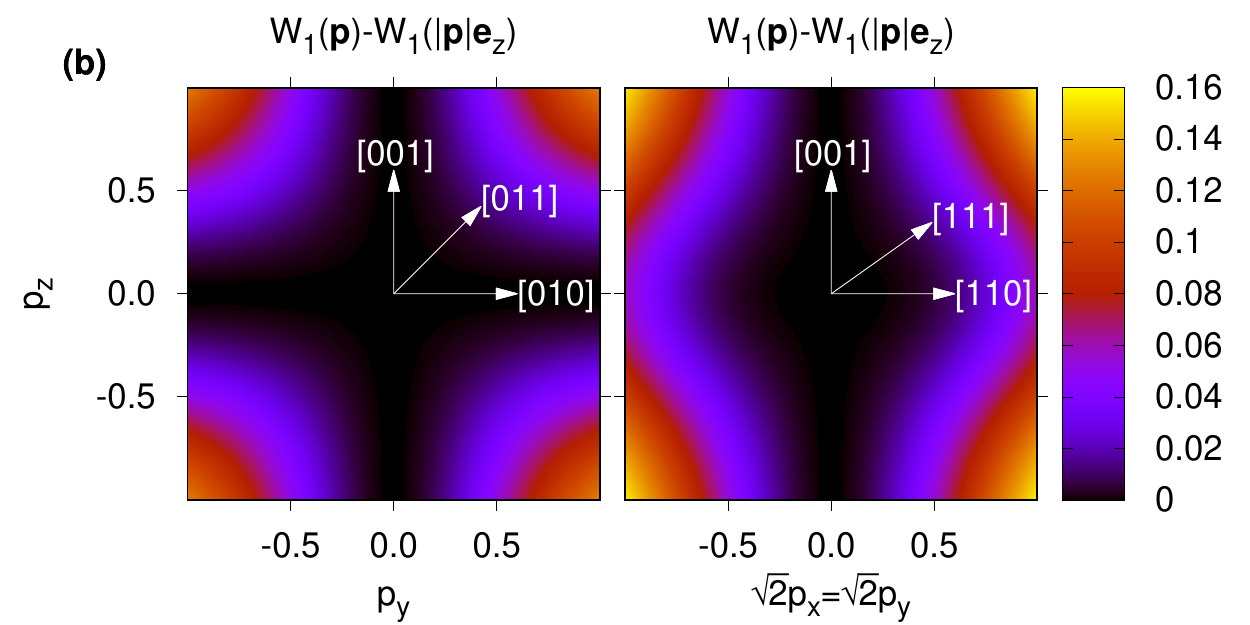}
  \caption{(a) Energy surfaces related to the yellow and green exciton
    series along the [$001$] axis.  (b)  Energy surface
    $W_{1,2}(\boldsymbol{p})$ in the planes normal to the [$100$] (left) and
    [$1\bar 10$] axis (right), respectively.  For better visualization of the
    $C_{4 \rm v}$ and $C_{2 \rm v}$ symmetries the corresponding energy
    values along the $z$-axis $W_1(|\boldsymbol{p}|\boldsymbol{e}_z)$ 
    have been subtracted.
    \label{fig1:EnergySurface}}
\end{figure}

As the hydrogen-like part of the Hamiltonian does not depend upon the
spin quantum numbers, only the Hamiltonian~\eqref{eq:Hb} describing
the band structure must be considered for the spin dynamics.
The spin part of the wave function can be expanded in the basis
$\ket{m_I, m_{S_{\rm h}}}$ with $m_I$ and $m_{S_{\mathrm{h}}}$ the
magnetic quantum numbers of the quasispin and hole spin as
\begin{equation}
  \ket{\psi}= \sum_{\substack{m_I=0,\pm 1\\ m_{S_{\rm h}}=\pm 1/2}}
  c_{m_I, m_{S_{\rm h}}}(\boldsymbol{p}) \ket{m_I, m_{S_{\rm h}}} \, .
  \label{eq:ansatz}
\end{equation}
The coefficients $c_{m_I, m_{S_{\rm h}}}$ depend on the classical
momentum $\boldsymbol{p}$ of the exciton.
Using the ansatz \eqref{eq:ansatz} in the Schrödinger equation
$H_{\mathrm{b}}(\boldsymbol{p},\boldsymbol{I},\boldsymbol{S}_{\rm h})
\ket{\psi}=W \ket{\psi}$
and multiplying from the left with $\ket{m'_I, m'_{S_{\rm h}}}$
leads to a six-dimensional Hermitian eigenvalue problem of
the form
\begin{equation}
  \boldsymbol{H}_{\mathrm{b}}\boldsymbol{c}=W \boldsymbol{c} \, ,
\end{equation}
which can be numerically solved by an appropriate \textit{LAPACK}
routine \cite{lapackuserguide3}.
A similar procedure was used for band-structure calculations in
Refs.~\cite{Schoene2016,SchoeneLuttinger}.
The eigenvectors $\boldsymbol{c}$ contain the coefficients from
Eq.~\eqref{eq:ansatz}. 
The six eigenvalues $W$ yield three distinct energy surfaces
$W_k(\boldsymbol{p})$ that are two-fold degenerate due to Kramers
theorem for half-integer spin systems with time-reversal symmetry.
The energies along the [$001$] axis are shown in
Fig.~\ref{fig1:EnergySurface}(a).
In this figure and in the following all parameters are given in
exciton-Hartree units which are obtained by setting $\hbar = e
= m_{\mathrm{0}} / \gamma_1' = 1 / (4\pi\varepsilon_{\mathrm{0}} \varepsilon) = 1$.

When the quasispin and the hole spin are coupled to
$\boldsymbol{J} = \boldsymbol{I} + \boldsymbol{S}_{\mathrm{h}}$,
the two energetically lowest surfaces $W_{1,2}(\boldsymbol{p})$ can be
assigned to the approximate quantum number $J = 1/2$ associated with
the yellow exciton series.
The remaining energy surfaces can be assigned to $J = 3/2$
associated with the green exciton series \cite{Rommel2020}.
The energy surfaces recover the cubic $O_{\rm h}$ symmetry of the
system, i.e.,
\begin{equation}
  W_k (O_{\rm h}\boldsymbol{p})= W_k (\boldsymbol{p}) \, .
\end{equation}
Cuprous oxide has two distinct classes of mirror planes.
In the three planes normal to the [$001$] axis and its equivalents
a $C_{4\rm v}$ symmetry can be recovered.
The six planes normal to the [$1\bar 10$] axis and its equivalents
share the $C_{2\rm v}$ symmetry.
The  energy surfaces $W_{1,2}(\boldsymbol{p})$ for the yellow
exciton series in the two distinct 
mirror planes are presented in Fig.~\ref{fig1:EnergySurface}(b).
For a better visualization of their symmetry properties the
energy values $W_{1,2}(|\boldsymbol{p}|\boldsymbol{e}_z)$ along the
$z$ axis [see Fig.~\ref{fig1:EnergySurface}(a)] have been subtracted.
Evidently, the $C_{4\rm v}$ and the $C_{2\rm v}$ symmetries of the
corresponding planes in the crystal are recovered in the energy
surfaces.

Replacing the term
$H_{\mathrm{b}}(\boldsymbol{p},\boldsymbol{I},\boldsymbol{S}_{\rm h})$
in Eq.~\eqref{eq:H} with one of the energy surfaces $W_k(\boldsymbol{p})$
corresponding to the yellow or green exciton series yields the Hamiltonian
\begin{equation}
  \mathcal{H} = E_{\mathrm{g}} + \frac{\gamma'_1}{2m_0} \boldsymbol{p}^2
  -\frac{e^2}{4\pi\varepsilon_0\varepsilon|\boldsymbol{r}|}
  + W_k(\boldsymbol{p}) \, ,
\end{equation}
which only depends on the exciton coordinates $\boldsymbol{r}$ and
momenta $\boldsymbol{p}$, and can therefore be utilized to calculate
classical exciton orbits described by Hamilton's equations of motion
\begin{equation}
  \dot r_i = \frac{\gamma'_1}{2m_0} p_i + \frac{\partial
              W_k(\boldsymbol{p})}{\partial p_i} \; , \quad
  \dot p_i = -\frac{e^2}{4\pi\varepsilon_0\varepsilon}
              \frac{r_i}{|\boldsymbol{r}|^3} \, . 
\end{equation}

Since the symmetry properties of the crystal are transcribed to the
energy surfaces $W_k(\boldsymbol{p})$, classical orbits starting in
one of the symmetry planes stay in that plane forever.
In the following we focus on the yellow exciton series.
The exciton orbits are calculated using algorithms for solving
ordinary differential equations \cite{press1992numericalpadeapprox}
and numerical root search \cite{more1980user}.
The investigation of the classical exciton dynamics in the
two-dimensional symmetry planes shows that the dynamics in the
$(x=0)$-plane perpendicular to the [$100$] axis is stable, i.e., orbits
in that plane are robust against small perturbations of the $x$
coordinate and the corresponding momentum.
By contrast, the dynamics in the symmetry plane perpendicular to the
[$1\bar{1}0$] axis is unstable against perturbations out of the plane.
In the following, we investigate the exciton dynamics in the stable
$(x=0)$-plane.
The phase space structures can be visualized via a \ac{PSOS}, where
the phase-space coordinates $(y,p_y)$ are plotted whenever the
trajectory crosses the $y$ axis, i.e., $z=0$.
\begin{figure}
  \includegraphics[width=\columnwidth]{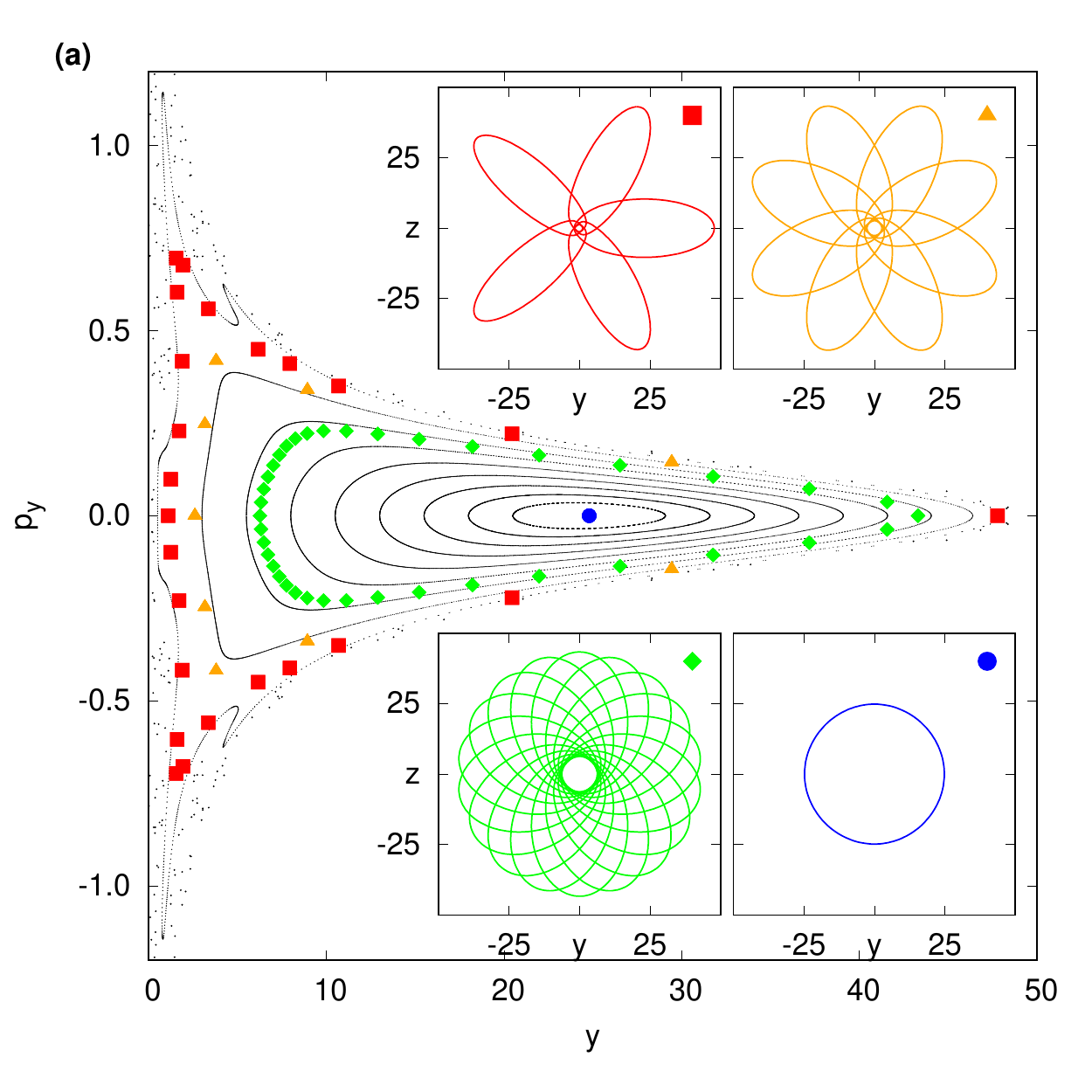}
  \includegraphics[width=\columnwidth]{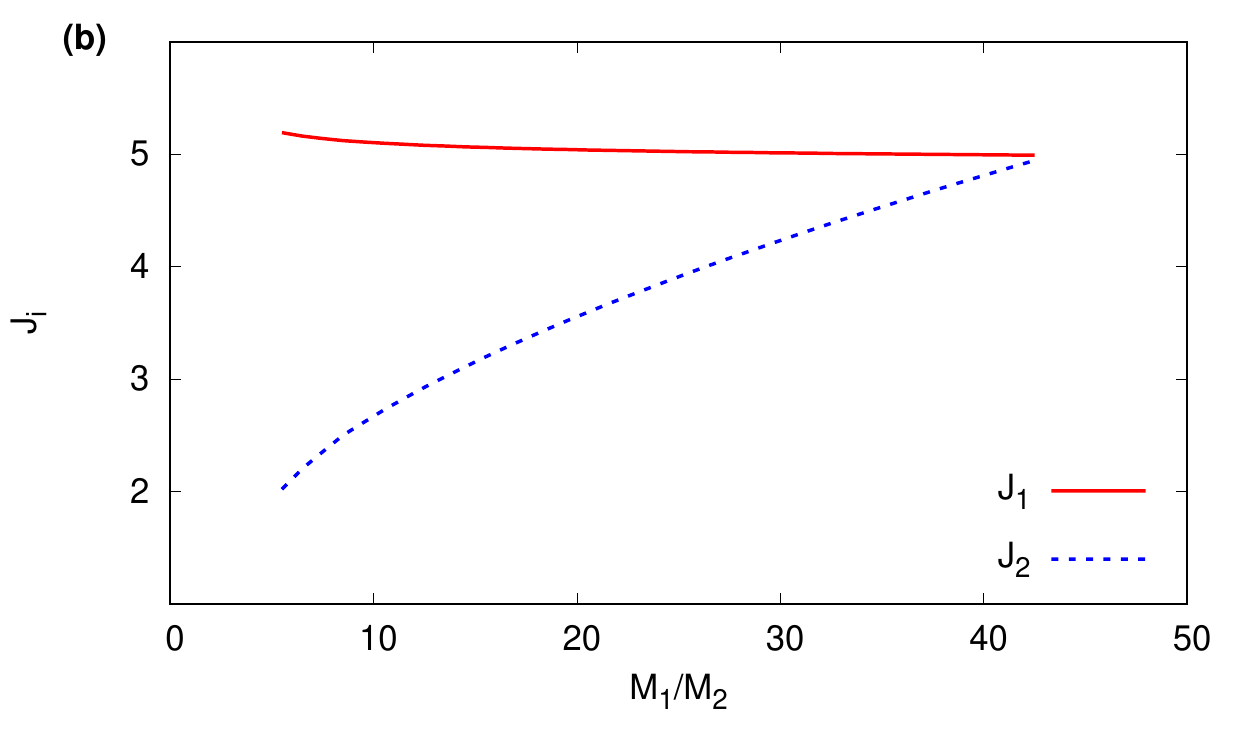}
  \caption{(a) \acf{PSOS} of the orbits in the $(x=0)$-plane at
    $n_{\mathrm{eff}}=5$.  Dominating regular tori are surrounded by a
    small chaotic region.  The insets show selected periodic orbits on
    rational tori marked with colored symbols.  (b) Action variables
    $J_1$ and $J_2$ as functions of the ratio $M_1/M_2$ of the
    rotation numbers.
     \label{fig2:psos}}
\end{figure}
The \ac{PSOS} at energy $E-E_{\mathrm{g}} = -0.02$ (in exciton-Hartree
units), which is related to an effective principal quantum number
$n_{\mathrm{eff}} \equiv [2(E_{\mathrm{g}}-E)]^{-1/2} = 5$, is presented in
Fig.~\ref{fig2:psos}(a).
This energy lies, on the one hand, in a range with $n_{\mathrm{eff}} > 3$
where the adiabatic approach discussed above is justified.
On the other hand, quantum theoretical calculations are still feasible
in this energy regime, and allow for the direct comparison of the
quantum and semiclassical calculations.
The \ac{PSOS} shows an elliptical fixed point at
about $y=24.79$, $p_y=0$ [blue dot in Fig.~\ref{fig2:psos}(a)], which
is surrounded by near-integrable tori.
Only in the outermost part of the \ac{PSOS} the torus structures break
down, i.e., the points in this area are distributed stochastically.
This indicates the onset of chaotic exciton dynamics
in the overall mixed regular-chaotic phase space.
Some selected rational tori are marked with colored symbols and the
corresponding periodic orbits are shown as insets in
Fig.~\ref{fig2:psos}(a).
Qualitatively similar structures as for $n_{\mathrm{eff}} = 5$ can be
observed when the energy is varied.
This will be discussed more detailed elsewhere.

The periodic orbits on the rational tori can be characterized by two
integer rotation numbers $(M_1,M_2)$, where $M_1$ describes the number
of Rydberg cycles of the orbit and $M_2$ is the number of rotations of
these cycles around the $x$ axis.
The classical action of the periodic orbits is given as
\begin{equation}
  S_{M_1,M_2} = 2\pi M_1 J_1 + 2\pi M_2 J_2 \, ,
\label{eq:S_M1M2}
\end{equation}
where $J_1$ and $J_2$ are the action variables of the torus, which
smoothly depend on the energy and the ratio $M_1/M_2$ of the rotation
numbers.
The ratio $M_1/M_2$ increases from the outermost tori towards the
central fixed point of the \ac{PSOS}.
This is related to a growing angular momentum of the classical orbits,
i.e., orbits in the outermost (chaotic) part of the \ac{PSOS} have low
angular momenta and the nearly circular orbit belonging to the central
fixed point has the highest angular momentum.

Using Eq.~\eqref{eq:S_M1M2} and its derivative
$dS_{M_1,M_2}/dM_1=2\pi J_1$, which can be well approximated by a
differential quotient for two consecutive periodic orbits with
rotation numbers $M_1$ differing by $1$, the action variables $J_1$
and $J_2$ can be constructed for the rational tori and interpolated
for the irrational tori.
The action variables $J_1$ and $J_2$ at $n_{\mathrm{eff}}=5$ as
function of the ratio $M_1/M_2$ are presented in Fig.~\ref{fig2:psos}(b).
They behave monotonically in the finite region between
$M_1/M_2\approx 5.5$, which is close to the border of the regular tori
[see Fig.~\ref{fig2:psos}(a)] and $M_1/M_2\approx 43.5$ related to the
central elliptical fixed point in the \ac{PSOS}.
For the nearly circular periodic orbit the two action variables
coincide, i.e., it is $J_1=J_2$ at the fixed point.

The action variables can in principle be used for a torus quantization
to obtain semiclassical eigenvalues of the excitons.
This, however, requires the extension of the two-dimensional torus
discussed above to a three-dimensional torus with action variables
$(J_1, J_2, J_3)$ including the orbits out of the symmetry planes.
Here we restrict ourselves to the quantization of the action variable
$J_1$, which then yields energy regions, where excitonic states with
a given principal quantum number $n$ can occur.
As can be seen in Fig.~\ref{fig2:psos}(b), for fixed $n_{\mathrm{eff}}$
the values of the action variable $J_1$ related to allowed classical
tori are restricted to a small and finite interval.
The ratio of the rotation numbers connected to the outermost 
regular torus stays approximately at $M_1/M_2=5.5$ when increasing 
$n_{\mathrm{eff}}$, whereas the ratio connected to the elliptical
fixed point $(M_1/M_2)_{\max}$ increases with increasing $n_{\mathrm{eff}}$.
\begin{figure}
  \includegraphics[width=\columnwidth]{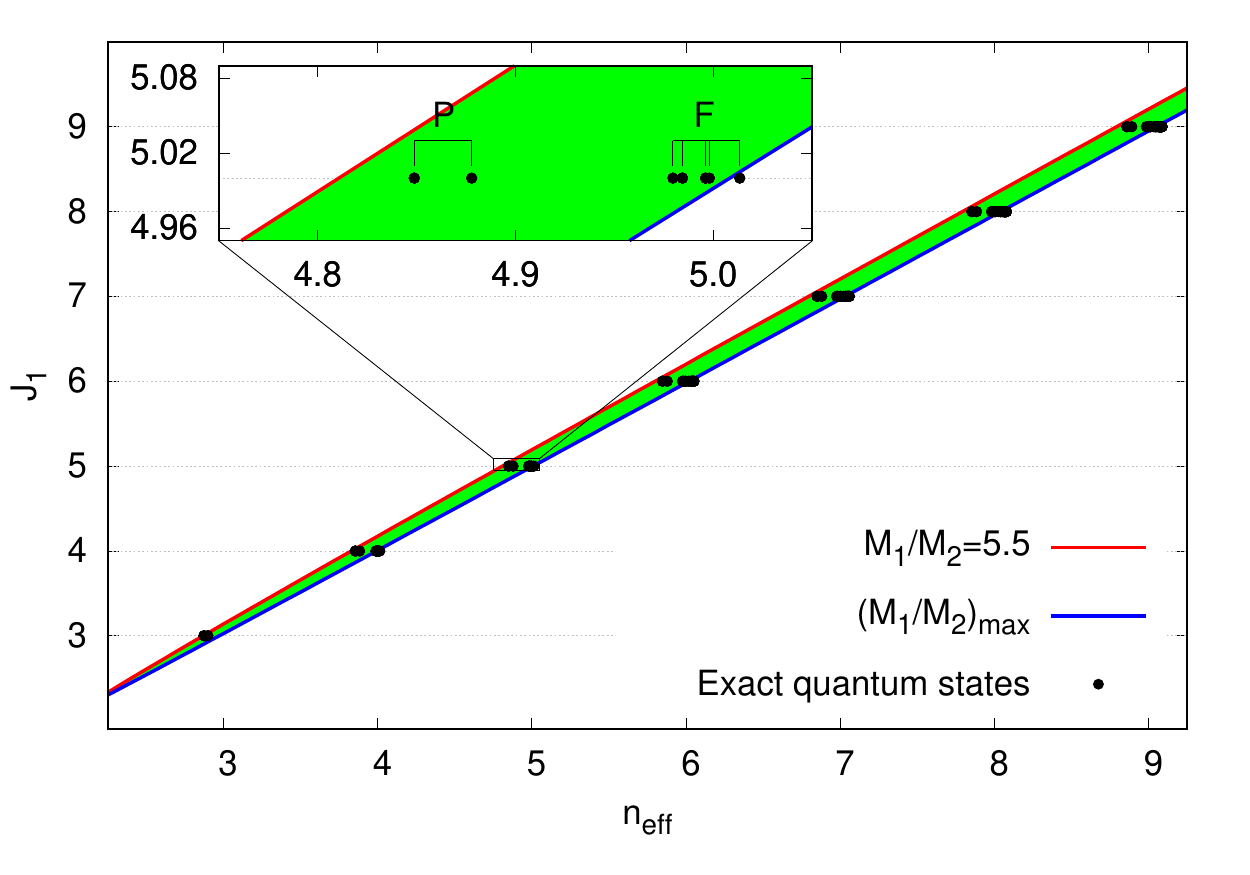}
  \caption{Action variable $J_1$ as function of the effective quantum
    number $n_{\mathrm{eff}}$.  The green area marks the allowed
    region of tori with ratio $M_1/M_2$ of the rotation numbers
    between $5.5$ (close to the outermost regular torus in the \ac{PSOS})
    and its maximum value $(M_1/M_2)_{\max}$
    (corresponding to the elliptical fixed point at the center of
    the \ac{PSOS}).  Semiclassically quantized exciton states belong to
    integer values of $J_1$.  The numerically exact quantum odd parity states with
    approximate principal quantum number $n$ are marked by black dots.
    In the inset the region around $n=5$ is enlarged and excitons are
    labeled as P (for $l=1$) or F (for $l=3$) states
    \cite{ObservationHighAngularMomentumExcitons,frankevenexcitonseries}.
    The black dots are perfectly located on
    the intersections of the lines $J_1=n$ with the green area of
    classical tori, which indicates the validity of the torus
    quantization for excitons.
    \label{fig3:Comparison}}
\end{figure}

In Fig.~\ref{fig3:Comparison} the $J_1$ intervals related to allowed
classical tori are shown as a function of the effective quantum number
$n_{\mathrm{eff}}$.
The green area marks the region of tori with ratio
$5.5 < M_1/M_2 \le (M_1/M_2)_{\max}$
of the rotation numbers, which cover nearly all classical orbits
except for those with very low angular momentum located in the
outermost (chaotic) region of the \ac{PSOS} [see Fig.~\ref{fig2:psos}(a)].
Semiclassical torus quantization requires $J_1=n$ with $n=1,2,3,\dots$
the principal quantum number.
The intersection of these lines with the green area in
Fig.~\ref{fig3:Comparison} now yields the allowed regions for the
fine-structure splitting of the $n$-manifolds, i.e., the intervals for
the effective quantum number (and thus the energy), where exciton
states with a given principal quantum number can exist.

To confirm the validity of the torus quantization we have computed the
fine-structure splitting of excitons in cuprous oxide by diagonalization
of the Hamiltonian using a complete basis set \cite{ImpactValence}.
We restrict the calculations to the odd parity states because the even
exciton series is dipole forbidden and cannot be observed in single
photon experiments
\cite{GiantRydbergExcitons,ObservationHighAngularMomentumExcitons}.
Furthermore, the even excitons require the consideration of
central-cell corrections \cite{frankevenexcitonseries}, which are not
included in our semiclassical approach.
The exact exciton states of the odd parity $n$-manifolds with $n=3$ to
$9$ are marked in Fig.~\ref{fig3:Comparison} by black dots on the
corresponding lines $J_1=n$.
They are perfectly located on the intersections of the lines $J_1=n$
with the green area of classical tori.
This clearly indicates the validity of the torus quantization for
excitons.
Approximate principal quantum numbers $n$ and angular momentum quantum
numbers $L$ can be assigned to each exciton state of the yellow series
\cite{frankevenexcitonseries}, for $n=5$ the dots are labeled as P or
F excitons in the inset of Fig.~\ref{fig3:Comparison}.
The F excitons appear at higher values of $n_{\mathrm{eff}}$ than the
P excitons, in accordance with the analysis of the classical exciton
dynamics as discussed above.
The agreement of the semiclassical and quantum mechanical results
in Fig.~\ref{fig3:Comparison} is remarkable when considering the
approximations, which have been used, i.e., the adiabatic approach for
the quasispin and hole spin dynamics, the semiclassical approximation,
and the restriction of the classical exciton dynamics to one of the
two-dimensional symmetry planes of the crystal.

%\paragraph{Conclusion}
In summary, we have investigated the classical exciton dynamics of
cuprous oxide.
The band structure of the crystal is taken into account via energy
surfaces in momentum space, which are obtained using an adiabatic
approach for the quasispin and hole spin.
Orbits in the plane perpendicular to the [100] axis exhibit a
quasi-periodic secular motion of Kepler ellipses on near-integrable
tori in phase space.
Semiclassical torus quantization allows for the interpretation of the
fine-structure splitting of $n$-manifolds in terms of the classical
exciton orbits.
The obtained results thus lead to a deeper and more comprehensive
physical understanding of excitons in semiconductors.
Future work will include, e.g., the semiclassical computation of
excitons based on the torus quantization of three-dimensional orbits,
and the extension of the semiclassical approach to challenging energy
regions of magnetoexcitons \cite{frankmagnetoexcitonscuprousoxide},
where numerically exact computations are no longer feasible.

\acknowledgments
This work was supported by Deutsche Forschungsgemeinschaft (DFG)
through Grant No.~MA1639/13-1 and through the International
Collaborative Research Centre (ICRC) TRR 160 (project A8). 
We thank Frank Schweiner for his contributions.

%\bibliography{paper}
%merlin.mbs apsrev4-1.bst 2010-07-25 4.21a (PWD, AO, DPC) hacked
%Control: key (0)
%Control: author (8) initials jnrlst
%Control: editor formatted (1) identically to author
%Control: production of article title (-1) disabled
%Control: page (0) single
%Control: year (1) truncated
%Control: production of eprint (0) enabled
%

\end{document}